# Dynamic load balancing algorithm of distributed systems

Lyudmila Kirichenko, Igor Ivanisenko, Tamara Radivilova

*Abstract* – **The dynamic load balancing algorithm based on the monitoring server load, self-similar characteristics of passing traffic have to provide a statistically uniform load distribution on servers, high performance, fault tolerance and capacity, low response time, the amount of overhead and losses was propose in work. Integrated measurement for the total imbalance level of the system were entered.**

## I. INTRODUCTION

Significant weakly predictable load fluctuations are typical for modern communication networks [1,2]. Similar conditions make inappropriate using of static planning strategies and network management which are acceptable in traditional telephony. In modern traffic engineering systems there are two approaches to management of network, which are known online- and offline-methods [3]. The main difference between these approaches is the timeline, which deals with the behavior of the matrix load, describing the transmitted traffic [1,3,4]. The offline-control system predicts long-term changes in the matrix load, while online-methods monitor the current state of the network in near real time. Under that the optimal scheme of load balancing can be developed. The main disadvantage of offline-control system is impossibility of respond to the current fluctuations in the load matrix which do not match the developed forecast. It should be noted that to date offline-approach is used in most traffic management systems.

Methods online traffic control operate usually at minute time intervals, do not use forecasting and able to cope with random fluctuations in load due to the possibility of obtaining information about the current state of the load matrix. The important difference these methods is mandatory autonomy, that eliminates operator intervention. As drawbacks online-approaches the lack of global forecasting of behavior of the load matrix and the resulting instability of routes caused by current fluctuations of traffic intensity are often called. A possible solution of these problems is the development of adaptive method of combining these approaches, including monitoring the current situation and using of forecasting in order to reduce the frequency of re-routing [1,3,4].

Traffic today's global network shows the exponential growth, there is significant structural changes, it is increasingly becoming difficult to predict fluctuations in load, weakly predictable load fluctuations become more tangible. The processes of networks convergence have led to the dominance of the IP protocol as a universal for all types of transmission data. However, the lack of built-in mechanisms of traffic engineering raises questions about the need to develop methods that allow more efficient use of existing network infrastructure, but does not require changing fundamentals of the global network. One of the most perspective solutions of these problems today are decentralized mechanisms balancing traffic.

Numerous studies of processes in information networks have shown that network traffic has the property of scale invariance (self-similarity). Self-similar traffic has a special structure, conserved on many scales – there are large amount of bursts with a relatively small average level of traffic. These bursts cause significant delay and packet loss, even when the total load of all flows is far from the maximum permissible values.

Self-similar properties discovered in the local and global networks, particularly traffic Ethernet, ATM, applications TCP, IP, VoIP and video streams. The reason for this effect lies in the features of the distribution of files on servers, their sizes, the typical behavior of users.

Now the multifractal properties of traffic are intensively studied. Multifractal traffic is defined as an extension of self-similar traffic due to take account of properties of second and higher statistics.

The purpose of this work is the development of dynamic load-balancing algorithm to the packet communication network which reduces the amount of overhead, based on the monitoring server load of distributed system, taking into account the fractal traffic structure.

## II. THE SYSTEM OF LOAD BALANCING

Information system can be a distributed network in structure, which is composed of many servers, storage nodes, and network devices. Each node is formed using a series of resources such as CPU, memory, network bandwidth, etc. Each resource has its own corresponding properties and there are many different types of resources. Described information system consists of group of servers and a load balancer. Load balancing system is proposed to build on the basis of subsystems

Ludmila Kirichenko – Kharkiv National University of Radioelectronics, av.Science, 14, Kharkiv, 61166, UKRAINE, E-mail: ludmila.kirichenko@gmail.com

Igor Ivanisenko - – Kharkiv National University of Radioelectronics, av.Science, 14, Kharkiv, 61166, UKRAINE, E-mail: ivanisenko79@yahoo.com

Tamara Radivilova - Kharkiv National University of Radioelectronics, av.Science, 14, Kharkiv, 61166, UKRAINE, E-mail: tamara.radivilova@gmail.com

of load balancer and control and monitoring that cooperate with each other closely:

- load balancer subsystem: load balancing algorithm, information about the current state of the system, flexible configuration QoS, dynamic allocation of traffic over various communication channels and nodes based on their current state, loading extend, the administrative load-balancing policies.

- control and monitoring subsystem: to collect and analyze statistics about current state of the system, finding the multifractal properties of the incoming data stream, the calculation of flow distribution in the network nodes based on traffic classification and utilization of servers and communication channels.

The load balancer $LB$ received traffic of intensity $\lambda = [\lambda_1, \lambda_2, ..., \lambda_N]$ at each moment $t \in T$. Traffic belong to the $qs$-th class of service which must be delivered to a server $Serv_k$ for processing, without exceeding defined maximum permitted delay values $\tau_{qs}$ and maximum percentage of losses $l_{qs}$ depending on current servers load and the actual channel capacity at moment of time.

Traffic has a lot of characteristics $V = \{\lambda, h, \mu\}$, where $\lambda = [\lambda_1, \lambda_2, ..., \lambda_N]$ are several independent multifractal applications flows (packages) with varying intensity; $h = [H, h(q), \Delta h]$, where $h(q)$ is selective value of the function of the generalized Hurst exponent, $H = h(2)$ is the value of Hurst exponent, $\Delta h = h(q_{min}) - h(q_{max})$ is the range of values of the generalized Hurst exponent for the area of traffic, $\mu$ is the requirement of the request. The requirement of the request is defined as a vector of required resources $\mu = (CPU, Net, RAM)$ to complete the request. Each $qs$-th class of service corresponds to a set of vectors of requirement resources $\mu_r = (CPU, Net, RAM)$, $r = 1, 2, ...$.

The load balancer $LB$ and servers $Serv_k$ are connected together by bilateral communication links [1,2] with the maximum capacity $Link_i = \{L_i\}$, $i = 1, 2, ... N$, that have utilization of network bandwidth $Net_i(t) = \{Net_i\}$, at time $t$. Each server $Serv_k$, $k = 1, 2, ...$ has the following characteristics: $CPU_i(t)$ - is the amount of utilization of $i$-th server CPU at the time $t$, which is calculated as $CPU_i(t) = \sum_{j=1}^{Mj} Load_{Mj}$ where $Load_{Mj}$ - loading of each core $j = \{1, 2, ...\}$ of each server processor $M = \{1, 2, ...\}$ at a time $t$. That is the core of distributed processors - does not matter.

Distribution of cores of processors does not matter. Two quad-core processors are correspond to four dual-core and correspond to eight single-core processors. All that matters is the total amount of cores. If the average load continuously exceeds 0.70, it is necessary to find out the reason for such system behavior in order to prevent problems in the future. If the load average exceeds 1.00, it need to find the cause and fix it urgently. $RAM_i(t)$ - is the amount of utilization of $i$-th server memory at a time $t$.

The input $LB$ received several independent multifractal applications flows (packages) with varying intensity $\lambda = [\lambda_1, \lambda_2, ..., \lambda_N]$. Each flow is sent to a queue $Q_i$ of limited capacity. Queuing time is dependent on the class of service $qs$, i.e. the priority request is taken into account (the highest priority is the first). While all priority service requests will not be processed, packages of other types stay in queue until the end of their lifetime. Newly received priority requests dropping off the process non-priority ones and with a probability equal to one displace them in storage (if have free waiting space), or outside the system (if the storage is full). Packages displaced from service are join to queue of non-priority requirements and can be serviced after all the priority ones. Queues are separate, free space fully accessible for all newly received requests. Unlike typical priority queue system the considered system is equipped with probabilistic eject mechanism. Priority package, that found all places busy during of processing other priority packet, is replacing one of the lower priority packets from the queue with a given probability, and takes his place. Displaced package is lost or sent back to the queue. The load balancer $LB$, in accordance with a specified algorithm, extract tasks from queues $Q_i$ and assign them to available cores of suitable servers.

To describe the mechanism of extrication of the network resources that occupied by the traffic at the end of transmission of traffic $qs$-th class of service (this occurs on the basis of data received from the routing protocol that supports communication on the available bandwidth and the available resources at the server (e.g., CSPF, SNMP), let introduce the variable $\varepsilon_{Net_i}^{qs,t_0}(t) = \{0,1\}$ that indicating that at the time $t$ the traffic class $qs$ stopped coming to server ($\varepsilon = 1$), which was come to service at the time $t_0$ and had passed on the path $Net_i(t)$ to the server $Serv_k$.

This variable contains all necessary data to determine the network resources that have to be extricated.

Load Balancer $LB$ at moment $t$ t is characterized by coefficient of loss $X_{LB}^{qs}(t) \in X$, the average waiting time of package in the queue $T_{LB}^{qs}(t) \in T$. The variable $X_{LB}^{qs}(t) \in X$ is the percentage of loss traffic of class of service $qs$ at load balancer, that transmitted over the path $Net_i(t)$ to the server $Serv_k$ at the moment $t$. It is assumed that the probability of package error in the path can be neglected and losses occur only in balancer because buffer overflows.

Queues and losses, that generated by traffic with multifractal properties, depend on the characteristics of the multifractal: generalized Hurst exponent function. Value range of the generalized Hurst exponent corresponds to the power of heterogeneity of traffic, i.e. it characterizes the scatter of the data [5]. The value of the ordinary Hurst parameter, obtained from generalized index, corresponds to the power of long-term dependence of realization and characterizes the correlation properties of traffic.

Condition monitoring servers and a free throughput can be done in three ways [3]: after each received a request; at fixed time intervals determined by a static algorithm; at non-fixed time intervals determined by the dynamic algorithm.

The information obtained by the first method, is the largest size, since measurements are carried out after each entered request. By the second method the amount of information is constantly, but it is necessary to determine the interval of information reading for the amount of information that has not been redundant and insufficient. In the third method, the amount of information depends on the frequency control intervals which must adapt to the structure of the incoming traffic due to its self-similar structure. Thus, the usage of dynamically changing algorithm is the most appropriate in terms of reducing data redundancy.

In this way, the frequency of monitoring will depend on the values of the generalized Hurst exponent $\Delta h = h(q_{min}) - h(q_{max})$ of incoming flow. The monitoring interval should be more frequent with a given value of the generalized Hurst exponent of incoming flow.

## III. THE DYNAMIC LOAD BALANCING ALGORITHM

The paper proposes the following dynamic dependent on multifractal properties of incoming traffic the algorithm of load balancing. The value of forecast have to be changed depending on changes in the parameters of incoming flow. Step by step description of the dynamic load balancing algorithm:

1) in the traffic that input to the load balancer, select the window $X$ of the fixed length $T$;
2) find the selective value of the function of the generalized the Hurst exponent $h(q)$, the value of Hurst exponent $H = h(2)$ and the range of values of the generalized Hurst exponent $\Delta h = h(q_{min}) - h(q_{max})$ for the area of traffic in the dedicated window;
3) collect and analyze statistical data: the intensity of the incoming flows $\lambda_1, \lambda_2, ..., \lambda_N$, utilization of network bandwidth $Net_i(t)$, the amount of utilization of $i$-th server CPU $CPU_i(t)$, the amount of utilization of $i$-th server memory $RAM_i(t)$ at a time $t$;
4) on the basis of multifractal properties of traffic (values of p.2), and the traffic intensity calculate necessary amount of resources for each $qs$ traffic classes;
5) perform calculations of flow distribution of network nodes based on traffic classification and utilization of servers and communication channels. On the basis of the data obtained is forecasting workload of the servers in the next step;
6) balancing traffic across the servers, according to the load balancing algorithm considering each class of flows;
7) perform balancing of underestimating the forecasting number of resource $CPU_i(t), Net_i(t), RAM_i(t)$. Reassessment of is not considered an algorithm because it does not introduce any significant;
8) collect data about servers utilization $CPU_i(t), Net_i(t), RAM_i(t)$ and send it to the system load balancing to calculate the new flow distribution;
9) move the window $X$ of the fixed length $T$ forward by a predetermined shift amount $\Delta T$;
10) carry out the traffic analysis and forecast of the next value of server load.

The proposed load balancing algorithm turned to providing a statistically uniform distribution of the load on servers, high performance, capacity, fault tolerance (automatically detecting failures of nodes and redirecting the flow of data among the remaining) and low response time, the quantity of service information and losses.

The load balancing algorithm should allocate requests to the server so that the deviation of server load from the mean value was minimal.

Lets enter an integrated measurement for the total imbalance level of the system, as well as the average imbalance level of each server [6]. It is proposed integrated load balancing metric as follows:

$$B = \frac{aN1_i C_i}{N1_m C_m} + \frac{bN2_i M_i}{N2_m M_m} + \frac{cNet_i}{Net_m} \qquad (1)$$

The referred physical server $m$ is selected first. Then, other physical servers $i$ are compared to server $m$. $N1_i$ is the CPU capability, $N2_i$ is the memory capability. Here, $C_i$ and $M_i$ denote the average utilization of CPU and memory, respectively. $Net_i$ represents the network throughput. Here, a, b, c denote the weighting factors for CPU, memory and network bandwidth, respectively. The major idea of this algorithm is to select the smallest value B among all physical servers. This technique is converting 3D resource information into a 1D value.

Considering the advantages and disadvantages of existing metrics for resource scheduling, an integrated measurement for the total imbalance level of a system, as well as the average imbalance level of each server, has been developed for load balancing strategy. Other metrics for different scheduling strategies can be developed as well. The following parameters are considered:

1. Average CPU utilization $CPU_i^u$ of a single server $i$. This is defined as the averaged CPU utilization during an observed period.

2. Average utilization of all CPUs in a system. Let $CPU_i^n$ be the total number of CPUs of server $i$,

$$CPU_u^A = \frac{\sum_i^N CPU_i^u CPU_i^n}{\sum_i^N CPU_i^n} \quad (2)$$

where N is the total number of physical servers in a system. Similarly, the average utilization of memory, network bandwidth of server $i$, all memories, and all network bandwidth in a system can be defined as $RAM_i^u$; $Net_i^u$; $RAM_u^A$; and $Net_u^A$, respectively.

3. Integrated load imbalance value ($ILB_i$) of server $i$. Variance is widely used as a measure of how far a set of numbers are spread out from each other in statistics. Using variance, an integrated load imbalance value ($ILB_i$) of server $i$ is defined:

$$\frac{(Avg_i - CPU_u^A)^2 + (Avg_i - RAM_u^A)^2 + (Avg_i - Net_u^A)^2}{3} \quad (3)$$

where $Avg_i = \frac{CPU_i^u + RAM_i^u + Net_i^u}{3} \quad (4)$

($ILB_i$) is applied to indicate load imbalance level comparing utilization of CPU, memory, and network bandwidth of a single server itself.

4. The imbalance value of all CPUs, memories, and network bandwidth. Using variance, the imbalance value of all CPUs in a data center is defined as

$$IBL_{cpu} = \sum_i^N (CPU_i^u - CPU_u^A)^2 \quad (5)$$

Similarly, imbalance values of memory and network bandwidth can be calculated. Then total imbalance values of all servers in a system is given by

$$IBL_{tot} = \sum_i^N IBL_i \quad (6)$$

5. The average imbalance value of a physical server $i$ is defined as

$$IBL_{avg}^{PM} = \frac{IBL_{tot}}{N} \quad (7)$$

where $N$ is the total number of servers. As its name suggests, this value is used to measure imbalance level of all physical servers.

6. The average imbalance value of a system is defined as

$$IBL_{avg}^{sys} = \frac{IBL_{cpu} + IBL_{RAM} + IBL_{Net}}{N} \quad (8)$$

## IV. CONCLUSION

Described in this paper load balancing system built on the basis of subsystem load balancer and subsystem control and monitoring that closely interact with each other. On the basis of the developed load balancing system is provided a dinamic algorithm of multifractal load balancing. Dynamic load balancing algorithm is based on the monitoring server load, the characteristics of self-similar traffic and turned to providing a statistically uniform distribution of the load on servers, high performance, fault tolerance and capacity, low response time, the quantity of service information and losses. In the developed algorithm data redundancy is reduced by dynamically changing the frequency intervals of control that adapts to the structure of incoming traffic given its self-similar structure. In this algorithm, the number of classes of flows depends on resource requirements. Also metrics are introduced for the each server and total imbalance level of the system. In the further work is necessary to simulate this algorithm in a distributed system and experimental study of the effectiveness of the proposed algorithm in comparison with the existing by changing the parameters of the incoming traffic.

## REFERENCES

[1]. Tarasov V., Polezhaev P., Shukhman A., Ushakov Y., Konnov A. Mathematical models of cloud computing data center using Openflow. Vestnik of OSU. Vol. №9 (145). 2012. Pp.150-155.

[2]. Andre. Understanding Linux CPU load - when should you be worried? 2009. URL: http://blog.scoutapp.com/articles/2009/07/31/understanding-load-averages

[3] E.I.Ignatenko, V.I.Bessarab, I.V.Degtyarenko. An adaptive algorithm for monitoring network traffic cluster in the load balancer. Naukovi pratsi DonNTU. Vol.21(183). 2011. Pp.95-102.

[4] A.A. Dort-Goltz. Development and research of a method of balancing the traffic in packet communication networks / Thesis for Ph.D. // St. Petersburg State University of Telecommunications. 2014. P.168.

[5] Ivanisenko I., Kirichenko L., Radivilova T. Investigation of self-similar properties of additive data traffic. Proc. X international scientific and technical conference CSIT. September, 2015. Pp.169-171.

[6] Wenhong Tian, Yong Zhao. Optimized Cloud Resource Management and Scheduling: Theories and Practices. Morgan Kaufman. 2014. P.284.